\documentclass[fleqn,usenatbib]{rasti}

\usepackage{newtxtext,newtxmath}
\usepackage{relsize}
\usepackage{xcolor}
\usepackage{booktabs}

\usepackage[T1]{fontenc}
\usepackage{amsmath}
\DeclareRobustCommand{\VAN}[3]{#2}
\let\VANthebibliography\thebibliography
\def\thebibliography{\DeclareRobustCommand{\VAN}[3]{##3}\VANthebibliography}

\usepackage{graphicx}	
\usepackage{amsmath}	
\usepackage{xcolor}

\title[Ariel Data Challenge 2022]{ESA-Ariel Data Challenge NeurIPS 2022: Introduction to exo-atmospheric studies and presentation of the Atmospheric Big Challenge (ABC) Database}

\author[Changeat \& Yip (2023)]{
Quentin Changeat,$^{1,2}$\thanks{ESA Research Fellow}\thanks{Domain expert: quentin.changeat.18@ucl.ac.uk}
and Kai Hou Yip$^{2}$\thanks{ML expert: kai.yip.13@ucl.ac.uk}
\\
$^{1}$European Space Agency (ESA), ESA Office, Space Telescope Science Institute (STScI), 3700 San Martin Drive, Baltimore MD 21218, USA\\
$^{2}$Department of Physics and Astronomy, Gower St., London WC1E 6BT, United Kingdom\\
}

\date{Accepted 10/01/2023. Received 05/01/2023; in original form 29/06/2022}

\pubyear{2022}

\begin{document}
\label{firstpage}
\pagerange{\pageref{firstpage}--\pageref{lastpage}}
\maketitle

\begin{abstract}

This is an exciting era for exo-planetary exploration. The recently launched JWST, and other upcoming space missions such as Ariel, Twinkle and ELTs are set to bring fresh insights to the convoluted processes of planetary formation and evolution and its connections to atmospheric compositions. However, with new opportunities come new challenges. The field of exoplanet atmospheres is already struggling with the incoming volume and quality of data, and machine learning (ML) techniques lands itself as a promising alternative. Developing techniques of this kind is an inter-disciplinary task, one that requires domain knowledge of the field, access to relevant tools and expert insights on the capability and limitations of current ML models. These stringent requirements have so far limited the developments of ML in the field to a few isolated initiatives. In this paper, We present the Atmospheric Big Challenge Database (ABC Database), a carefully designed, organised and publicly available database dedicated to the study of the inverse problem in the context of exoplanetary studies. We have generated 105,887 forward models and 26,109 complementary posterior distributions generated with Nested Sampling algorithm.  Alongside with the database, this paper provides a jargon-free introduction to non-field experts interested to dive into the intricacy of atmospheric studies. This database forms the basis for a multitude of research directions, including, but not limited to, developing rapid inference techniques, benchmarking model performance and mitigating data drifts. A successful application of this database is demonstrated in the NeurIPS Ariel ML Data Challenge 2022.
\end{abstract}




\section{Context}

The field of exoplanet has come a long way since the discovery of the first exoplanet in 1994 \citep{Wolszczan_1992}. With the launch of dedicated telescopes for the detection of exoplanets, such as the Convection, Rotation et Transits planétaires \citep[CoRoT,][]{corot}, the Kepler \citep[][]{Borucki2010}, and the Transiting Exoplanet Survey Satellite \citep[TESS,][]{Ricker} space telescopes, we now have basic characteristics, such as planetary radii or masses, for more than 5000 alien worlds. From the observed population, we deduced that, while exoplanets are ubiquitous \citep{Cassan_2012, Batalha_2014}, the architecture of our solar system does not appear to be a typical outcome of planetary formation. For instance, the first detected exoplanet around a sun-like star is classified as a hot Jupiter \citep{Mayor_1995}, a planet of similar size to Jupiter (e.g about 10 times the size of Earth) but orbiting so close to its host-star that it completes a full revolution in about 4 days. Such planet does not exist in our solar-system and so are the majority of the observed planets, referred as sub-Neptunes due to their size being between the size of Earth and Neptune \citep{Howard_2010, Fulton_2017, Petigura_2022}. To answer the most fundamental questions of the field, such as "what are exoplanets made of?" or "how do planets form?", one must obtain complementary information to planetary masses and radii. 

In the last decade, astronomers have therefore turned their attention to exoplanetary atmospheres, or exo-atmospheres, in the quest for further constraints on these worlds \citep{Charbonneau_2002, Tinetti_2007, Swain_2008, Kreidberg_2014, Schwartz_2015, Sing_2016, Stevenson_2017, Hoeijmakers_2018, DeWit_2018, Tsiaras_2018, Tsiaras_2019, Brogi_2019, Welbanks_2019, Edwards_2020, Changeat_2021_k9,  Roudier_2021, Yip_2021, Changeat_2022, Evans_2022, Edwards_2022_pop, 2022_Estrela, 2022_Chen}. The study of exoplanet atmospheres has been enabled by the use of space-based instrumentation, such as the Hubble Space Telescope (HST), the retired Spitzer Space Telescope, and ground-based facilities such as the Very Large Telescope (VLT). Many discoveries were made. We, for instance, know that water vapour is present in many hot Jupiter atmospheres, and we have recently recovered evidence for links between atmospheric chemistry and formation pathways. However, with the recent launch of the NASA/ESA/CSA James Webb Space Telescope \cite[JWST][]{Greene_2016} and the upcoming ESA Ariel Mission \citep{Tinetti_2021_redbook} and BSSL Twinkle Mission \citep{Edwards_2019_twinkle}, the field of exoplanetary atmosphere will undergo a revolution. The quality and quantity of atmospheric data will be multiplied exponentially, bearing many new challenges.    

One of the main challenge in the study of exo-atmospheres, even today, concerns with the reliable extraction of information content from observed data. Atmospheres are complex dynamical systems, involving many physical processes (chemical and cloud reactions, energy transport, fluid dynamics), that are coupled, poorly understood, and difficult to reproduce on Earth. Astronomers have therefore attempted to interpret observations of atmospheres using retrieval techniques: simplified models (or reduced order models) for which the parameter space of possible solutions is explored using a statistical framework \citep{Irwin_2008, Madhusudhan_2009, Line_2012, Line_2013, Waldmann_2015, Waldmann_2015_2, Lavie_2017, Gandhi_2018, Molliere_2019, Zhang_2019, Min_2020, Al-Refaie_2021, Harrington_2022}. With current observational data, state-of-the-art retrieval models use sampling based Bayesian techniques, such as MCMC or Nested Sampling, with non-informative (uniform) priors to obtain the posterior distributions of between 10 and 30 free parameters \citep{Changeat_2021}. The number of free parameters depends on the information content available in the observational data and the chosen atmospheric model. As of today, there is no consensus on the most appropriate atmospheric model to employ, and we cannot obtain in-situ observations (e.g. we cannot travel there). Sampling-based techniques typically require between 10$^5$ and 10$^8$ forward model calls to reach convergence, meaning that only models providing spectra of the order of seconds are viable. With the increase in data quality, thanks to JWST, Ariel and Twinkle, it will enable a wider range of atmospheric processes to be probed by the observations, implying that forward models must grow in complexity and so does the dimensionality of the problem \citep{ERS_2022}. As such, interpreting next-generation telescope data is currently a real issue, which has been highlighted multiple times by studies relying on simulations, that will require a revolution in both our models and information extraction techniques \citep{Rocchetto_2016, Changeat_2019, Caldas_2019, Yip_2020_lc, Taylor_2020, Taylor_2021,  Changeat_2021, Al-Refaie_2021_t31, Yip_2022}. 

In recent years, the community started to explore alternative approaches to circumvent the bottleneck with sampling based approaches. Machine Learning models land itself as a promising candidate with its high flexibility and rapid inference time. \cite{Waldmann_2016} pioneered the use of deep learning network in the context of atmospheric retrieval, training a Deep Belief Network to identify molecules from simulated spectra. On the other end, \cite{Marquez2018} led the first attempt to train a Random Forest regressor to predict planetary parameters directly. Since then, the field has started to look at different ML methodologies to bypass the lengthy and computationally intensive retrieval process \citep{Zingales_2018,Soboczenski_2018,Cobb2019,Hayes2020,Oreshenko_2020,Nixon_2020,Himes2022,Ardevol2022,Haldemann2022,Yip_2022}. Pushed by astronomers' need for explainable solutions, other groups have also looked into the information content of exoplanetary spectra with AI \citep{Guzman-mesa2020, yip2021peeking}.  


The publicly available Atmospheric Big Challenge (ABC) Database of forward models and retrievals aims to provide the resources to address aforementioned issue via participation of external communities and encourage novel, cross-disciplinary solutions. It is constructed as a permanent data repository for further investigations. The database is accessible at the following link: \url{https://doi.org/10.5281/zenodo.6770103}. 

Since the creation of similar database constitutes a major barrier to anyone interested in applying Machine Learning in the domain of exoplanet atmospheres, we emphasise on its release as a community asset. The organisation and creation of this dataset poses a challenge on its own because:
\begin{enumerate}
    \item It requires a cross-disciplinary collaboration. The problem requires domain knowledge (atmospheric chemistry, radiative transfer, atmospheric retrievals) to ensure the data product represents a meaningful science case rather than a trivial example. At the same time, it requires machine learning expertise to ensure the data product is representative of the problem at hand, and ideally, one that adequately reflects the reality.
    \item It requires access to the relevant tools which is often exclusive to communities in exoplanet: atmospheric retrieval and chemistry codes as well as instrument noise simulators.
    \item It requires significant computing resources. For this project more than 2,000,000 CPUh were used. Simulations of this scale have never been attempted before.
\end{enumerate}

This paper is written to 1.) provide non-field experts with a light-weighted introduction to the science behind the data generation process,  2.) document the steps involved in the creation of the ABC database and 3.) To provide a carefully curated, well-organised, and scientifically relevant dataset for any research community. This manuscript complements the data challenge proposal description \citep{Yip_2022_neurips} accepted as a NeurIPS 2022 data challenge. It is intended to provide the required domain knowledge for non-field experts. We presents a simplified jargon-free introduction to the most commonly employed techniques in the field of exo-atmospheres in Appendix \ref{A0:intro}. 


\section{Data generation}

For the data generation we employed \emph{Alfnoor} \citep{Changeat_2021}, a tool built to expand the forward model and atmospheric retrieval capabilities of \emph{TauREx\,3} \citep{Al-Refaie_2021} to large populations of exo-atmospheres. \emph{Alfnoor} allows to automatise the generation or telescope simulations and perform large scale standardised atmospheric retrievals. A lightweight description of the main concepts behind atmospheric studies of exoplanets are described in Appendix \ref{A0:intro}. In the context of ESA-Ariel, we generated 105,887 simulated forward observations as well as 26,109 standardised retrieval outputs. 

\subsection{Source of input parameters}

\begin{figure}
 \includegraphics[width=0.95\columnwidth]{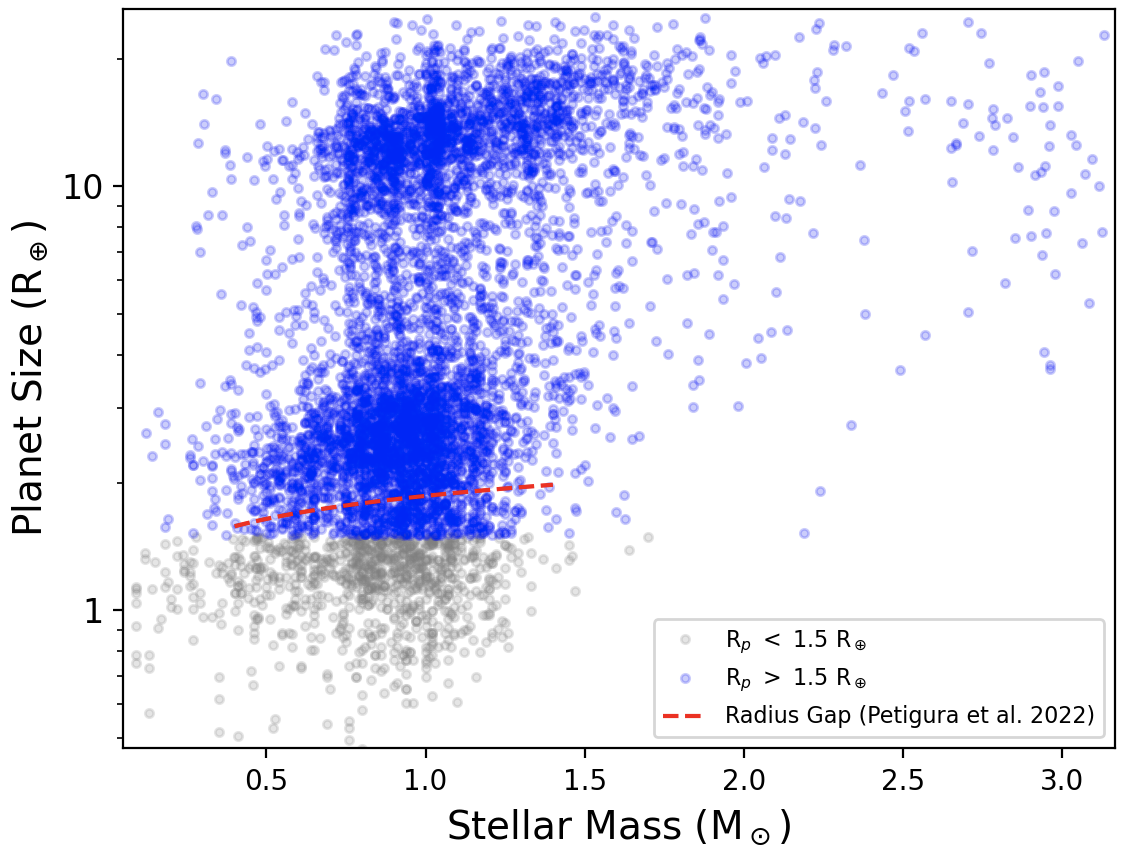}
 \caption{Size of the considered planets versus the Mass of their host star. We exclude planets that have radii below 1.5 R$_\oplus$ marked in Grey and approximately corresponding to the lower limit on the Radius valley.}
 \label{fig:mass_massstar}
\end{figure}

To model those extrasolar systems, some preliminary assumptions were required. In particular, all the parameters that are not linked to the atmospheric chemistry needed to be fixed to realistic values. Those parameters include, but are not limited to, stellar radius (R$_s$), distance to Earth (d), star magnitude K (Kmag), planetary radius (R$_p$), planetary mass (M$_p$), planet equilibrium temperature (T) and transit duration (t$_{14}$).

The planetary objects in this database were selected from the list of confirmed known exoplanets and the list of TESS exoplanet candidates (TOIs). This list was constructed as part of the ESA-Ariel Target list initiative \citep{Edwards_2019_TG1, Edwards_2022_TG2}, frozen to the 1st of March 2022 for this database. For the TOIs, we are aware that some of those objects will not be exoplanets, however the observation of their transit by TESS and the first preliminary checks of their inferred properties make them compelling objects. Follow-up observations will allow us to classify their nature, but for the purpose of building this database, they are as close as possible to what the reality looks like. As radial velocity follow-ups cannot and is not systematically conducted for all targets, the mass of some of those objects is unknown. In this case, as in \cite{Edwards_2022_TG2}, we replace the planetary mass by an estimate from the relation described in \cite{Chen_2017_mass}. To those lists of objects, we filtered all the planets with radius below 1.5 R$_\oplus$, the conservative value for the middle of the Radius Valley \citep{Fulton_2017, Cloudier_2020, Petigura_2022}. This is because the atmospheric composition of small planets would require a much more complex treatment (e.g the assumption of hydrogen dominated atmosphere is not theoretically sound) than is proposed here. In total, we obtained data for 2,972 confirmed exoplanets and 2,928 candidate exoplanets, thus bringing our total to 5,900 unique objects.

\begin{figure*}
    \centering
    \includegraphics[width=\textwidth]{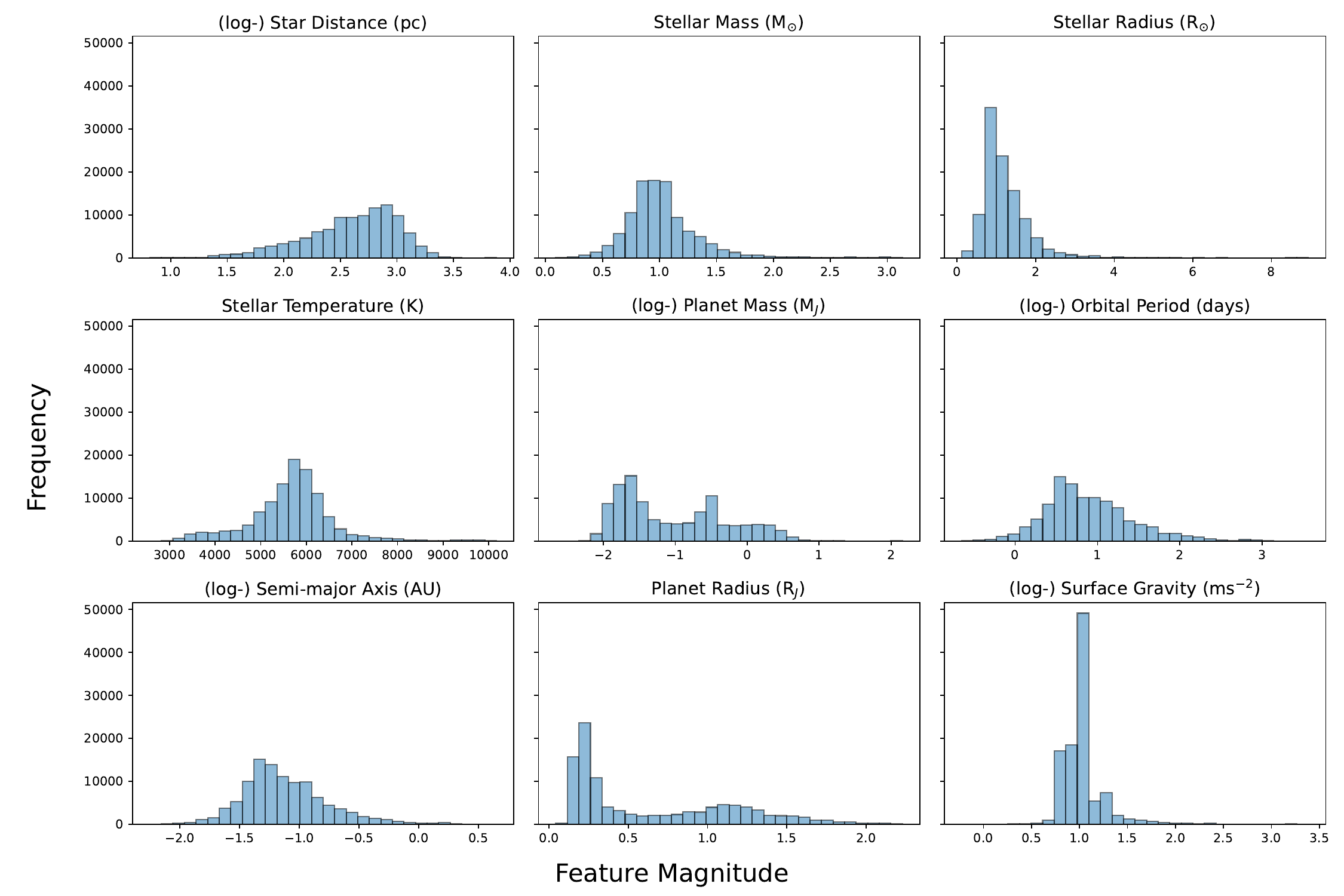}
    \caption{Distribution of nine stellar and planetary parameters used to generate the synthetic spectra. These distributions follow closely to the actual demographic of currently known population of exoplanets, and therefore they are also subject to biases presented in the original population.}
    \label{fig:supp_distribution}
\end{figure*}

Figure \ref{fig:supp_distribution} shows the distributions of 9 selected stellar and planetary parameters. These values are taken from the actual planetary system and therefore follows the current observed demographics, these values remains unchanged thorough the data generation process. However, relying on currently known planets is a double edged knife. While it saved us from making unverified assumptions, our data is prone to selection bias stemmed from the observation technique, strategy and instrument specification. These biases can be easily spot from Figure \ref{fig:supp_distribution}. For instance, the distribution of orbital period tends to be shorter (peaks around $\sim$3 days) as their proximity to the host star makes them easier to discover. 

\subsection{The atmospheric forward model setup}
 We produce batches of randomised observations for the population described in the previous section. For each planet the stellar parameters (R$_s$, d, Kmag), orbital (t$_{14}$) and bulk parameters (R$_p$, M$_p$, T) are fixed to their literature values, while the chemistry of the atmosphere is randomly generated. The thermal profile is assumed to be isothermal (constant temperature) at the equilibrium temperature of the planet, and we simulate the planet's atmosphere from 10 bar to 10$^{-10}$ bar using 100 layers (divided uniformly in log-pressure space).

For the chemistry, we assume a primary atmosphere made mainly from hydrogen and helium (He/H$_2$=0.17), to which we add trace gases. The trace gases are H$_2$O \citep{Polyansky_2018}, CH$_4$ \citep{Yurchenko_2017, Chubb_2021}, CO \citep{Li_2015}, CO$_2$ \citep{Yurchenko_2020} and NH$_3$ \citep{Coles_2019}, selected based on our current understanding of exoplanetary chemistry \citep{Agundez_2012, Venot_2015, Madhu_2016, Drummond_2016, Woitke_2018, Stock_2018, Venot_2020, Al-Refaie_2021_t31, Baeyens_2022}. The mixing ratio, or trace abundance, of those gases is randomly chosen using a Log Uniform law and depends on the molecule considered. The Log Uniform law is chosen rather than a more informative law (such as equilibrium chemistry) because we are looking for solutions that are unbiased to our current, most likely limited, understanding of atmospheric chemistry. Such training set is suitable to produce ML solutions behaving in a similar way to the so-called free chemistry retrievals. If correlation exist in a real population (for instance between the chemistry of the atmosphere and its thermal structure), such method should allow the extraction of this trend without the need to implicitly make a physical assumption. Note that this is required in the cases where data has undergone a data shift (in this case when the data is generated using a different atmospheric assumption). Another important point to consider involves the detection capabilities of Ariel for each molecule and the degeneracy between molecular species. For instance, CO shares similar features to CO$_2$ in Ariel but it is a much harder molecule to detect due to its weaker absorption properties. Due to those differences in strength of spectral features and guided by the Ariel Tier-2 detection limits investigated in \cite{Changeat_2020}, we select different bounds for the randomised chemical abundances. This process allows us to balance our dataset and ensure that a significant fraction of the planets have detectable amount of CO. The bounds employed for this dataset are:
\begin{enumerate}
    \item H2O: RandomLogUniform(min= -9, max= -3). 
    \item CO: RandomLogUniform(min= -6, max= -3). 
    \item CO2: RandomLogUniform(min= -9, max= -4). 
    \item CH4: RandomLogUniform(min= -9, max= -3). 
    \item NH3: RandomLogUniform(min= -9, max= -4). 
\end{enumerate}

For each parametrised atmospheres, we compute the radiative transfer (see Appendix A) layer-by-layer, including the contributions from molecular absorption, Collision Induced Absorption, and Rayleigh Scattering. 

Each spectrum is first computed at high-resolution\footnote{Spectra have to be computed at high-resolution ($\mathcal{R}$) since instrumental binning is done on the received photons, e.g the recorded transit depth $\Delta$. In our case, we used $\mathcal{R}$ $= \frac{\lambda}{\Delta \lambda} = 10,000$.}, before being convolved with an Ariel instrument simulation. For each planet, we employed the \emph{TauREx} plugin for \emph{ArielRad} \citep{Mugnai_2020_arielrad}, the official Ariel noise simulator, to estimate the noise on observation at each wavelength. With \emph{ArielRad}, we force each observation to satisfy the criteria for Ariel Tier 2 observations \citep{Tinetti_2021_redbook}, meaning that the observations have a specific resolution profile (e.g: R $\approx$ 10 for 1.10 $< \lambda <$ 1.95 $\mu$m; R $\approx$ 50 for 1.95 $< \lambda <$ 3.90 $\mu$m; R $\approx$ 15 for 3.90 $< \lambda <$ 7.80 $\mu$m) and that the signal-to-noise ratio (SNR) of the observations must be higher than 7 on average. The SNR is here defined on the atmospheric signal (e.g. the second part of Equation \ref{eq:deltatransit}). To produce the simulated spectra, we select the minimum number of transit that allow to reach this threshold, meaning that our sample of observations contains a wide range of final SNR. Since we used real objects for those simulations and that all planets are not favourable targets for Ariel, this means that some targets require an un-realistic number of observations to reach the SNR condition of Tier 2. However, this does not affect the purpose of this dataset, providing independent instances of realistic noise profiles.

\begin{figure*}
 \includegraphics[width=1.8\columnwidth]{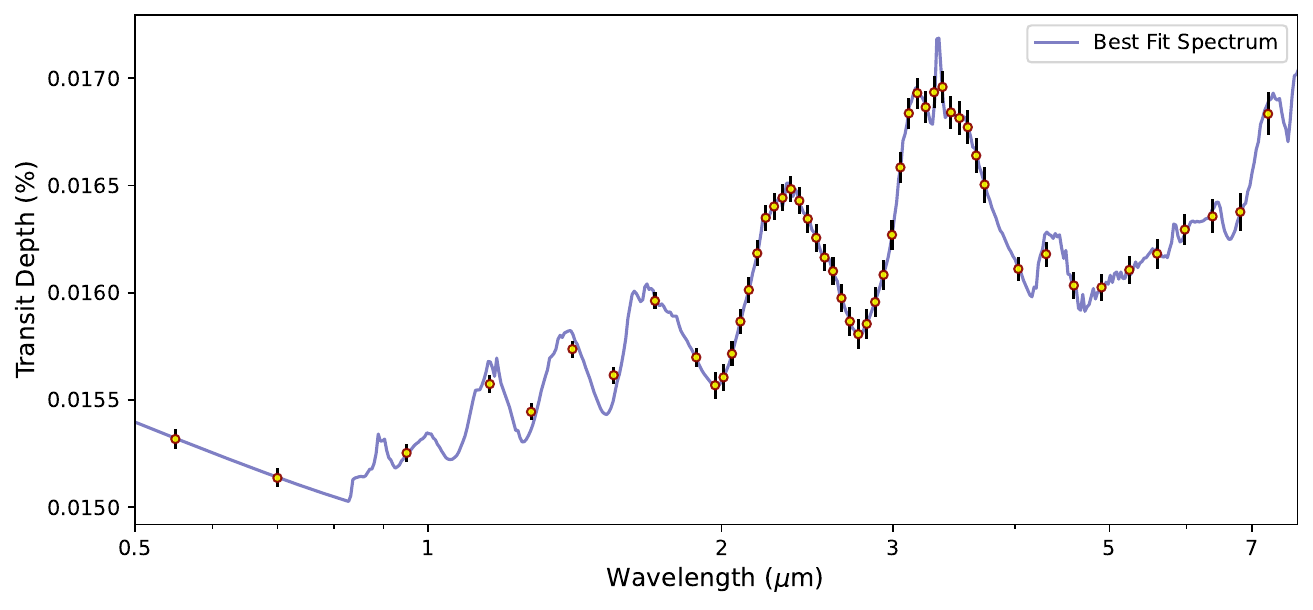}
 \caption{Example of a simulated Ariel observation with errorbars (datapoints) for a randomised chemistry. The best-fit model obtained using atmospheric retrieval is also shown (solid line). The slope at the lowest wavelengths arises from Rayleigh Scattering, while most of the other spectral modulations in this example can be attributed to CH$_4$. The datapoints around 4.5$\mu$m are associated with CO and CO$_2$ absorption. Note the difference in wavelength coverage (0.5$\mu$m to 7.8$\mu$m) as compare to the HST spectrum (1.1$\mu$m to 1.7$\mu$m) in Figure \ref{fig:sensitivity}, which allows us to extract precise information for many molecules.}
 \label{fig:retrieval_ariel}
\end{figure*}

Following those steps, we obtain a realistic Ariel simulated observation for each planet and each randomised chemistry. We show an example of such simulated observation in Figure \ref{fig:retrieval_ariel}. In total, we produced 105,887 simulated observations for the ABC Database.

\subsection{The atmospheric retrieval setup}
\label{sec:retrieval}

For 26,109 (25\%) of the simulated observations generated at the previous step, we perform the traditional inversion technique using \emph{Alfnoor}. 

For the model to optimise, we kept the same setup as presented in the previous section and performed parameter search on the following free parameters: isothermal temperature (T), log abundances for H$_2$O, CO$_2$, CH$_4$, CO and NH$_3$. The priors are made wide and un-informative, with the atmospheric temperature being fitted between 100K and 5500K and the chemical abundances between 10$^{-12}$ and 10$^{-1}$ in Volume Mixing Ratios. The widely used Nested Sampling Optimizer, MultiNest \citep{Feroz_2009}, was employed with 200 live points and an evidence tolerance of 0.5.

For a single example on Ariel data, we provide the best-fit spectrum in Figure \ref{fig:retrieval_ariel}. From the optimization process, we are able to extract the traces of each parameters and the weights of the corresponding models. This allows to construct the posterior distribution of the free parameters with, for instance \emph{corner}. The posterior distribution of the same example is shown in Appendix B, Figure \ref{fig:posteriors_ariel}. Processing of the posterior distribution also allows to extract statistical indicators describing the chemical properties of the planet, such as mean, median and quantiles for each of the investigated parameters.

\subsection{Data Overview}

\begin{figure}
    \centering
    \includegraphics[width=\columnwidth]{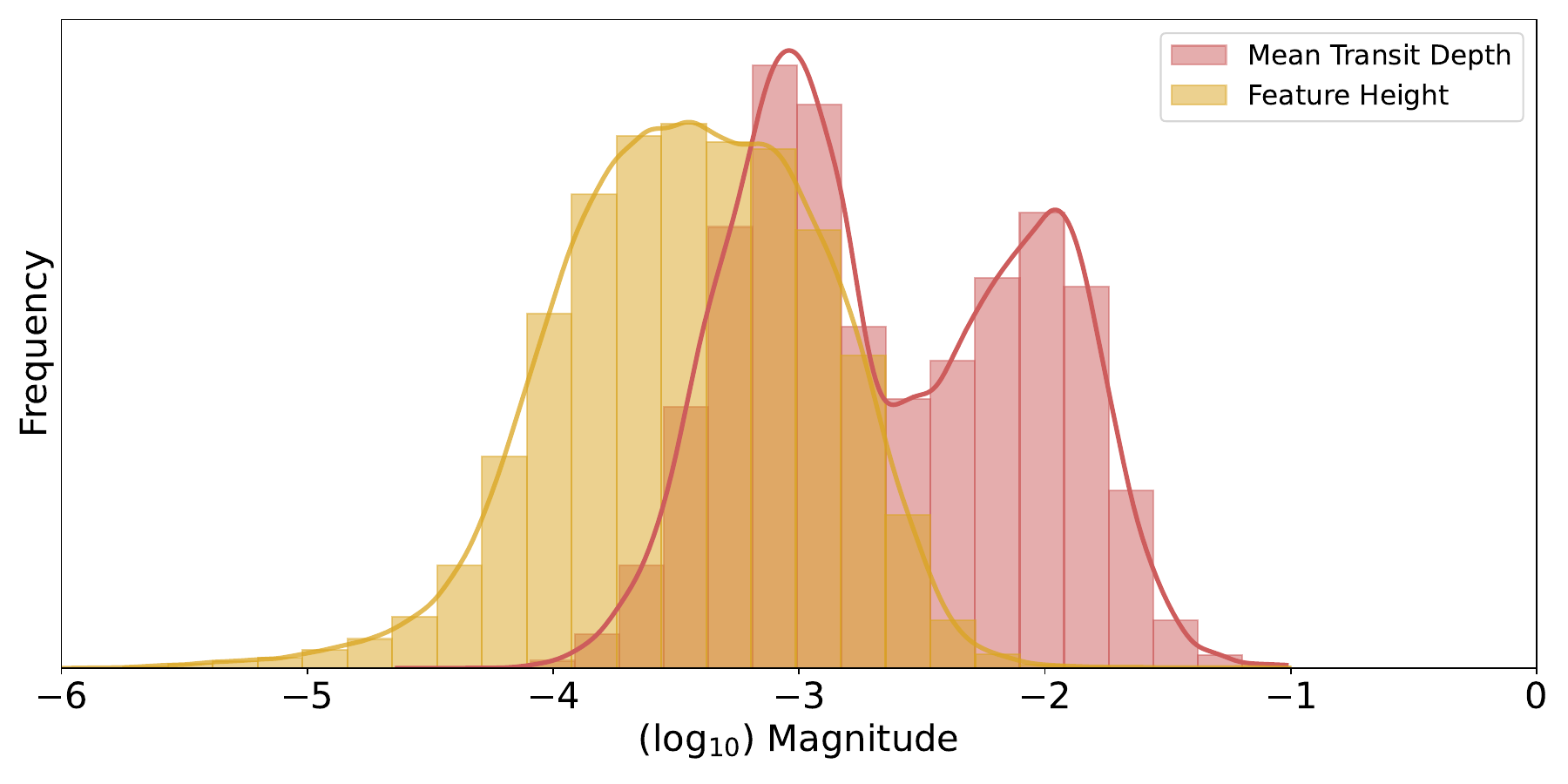}
    \caption{Distribution of mean transit depth (red) overlapped with the distribution of the feature height (orange), both measured in logarithm scale. The dichotomy displayed in mean transit depth distribution stemmed from the observational demographics of planet radius, showing the diversity of currently known exoplanets in our dataset. On the other hand, the feature height documents the ``strength" of spectroscopic features in each spectrum (such as absorption features or strong trends induced by Rayleigh scattering). Any successful model must be able to account for the variations in both scales.}
    \label{fig:two_distribution}
\end{figure}

Following the data generation process outlined above, we have generated a total of 105,887 forward models in Ariel Tier-2 resolution. 26\% of them are complemented with results from atmospheric retrieval (following a generic setting as described in Section \ref{sec:retrieval}). 

Figure \ref{fig:two_distribution} shows the distribution of mean transit depth (red) overlapped with the distribution of feature height (orange). The former served as a proxy of the diverse planetary classes present in the dataset. The characteristic dichotomy stemmed from current demographics studies\footnote{Latest studies show that Super-Earth sized planets are prevalent while there is a deficiency in the population of sub-Neptunes.} and selection bias in our observation technique \footnote{Transit technique tends to favour larger planets.}. The latter is calculated from the difference between the maximum and minimum transit depth of each spectrum, it served as a proxy of the ``strength" of the spectroscopic features presented in the spectra, e.g. the peaks and troughs as seen in Figure \ref{fig:sensitivity} and Figure \ref{fig:retrieval_ariel}. We note that an SED with linear slope will also produce a non-negligible feature height value, which is still considered as spectroscopic feature in our case. The two quantities are closely linked to our targets of interest, which means that any successfully model not only need to account for the inter-variation between different spectra , it also needs to take into account the intra-variation across wavelength channels, which is always 1-3 orders of magnitude smaller than the variation in mean transit depth.

\begin{figure*}
    \centering
    \includegraphics[width=\textwidth]{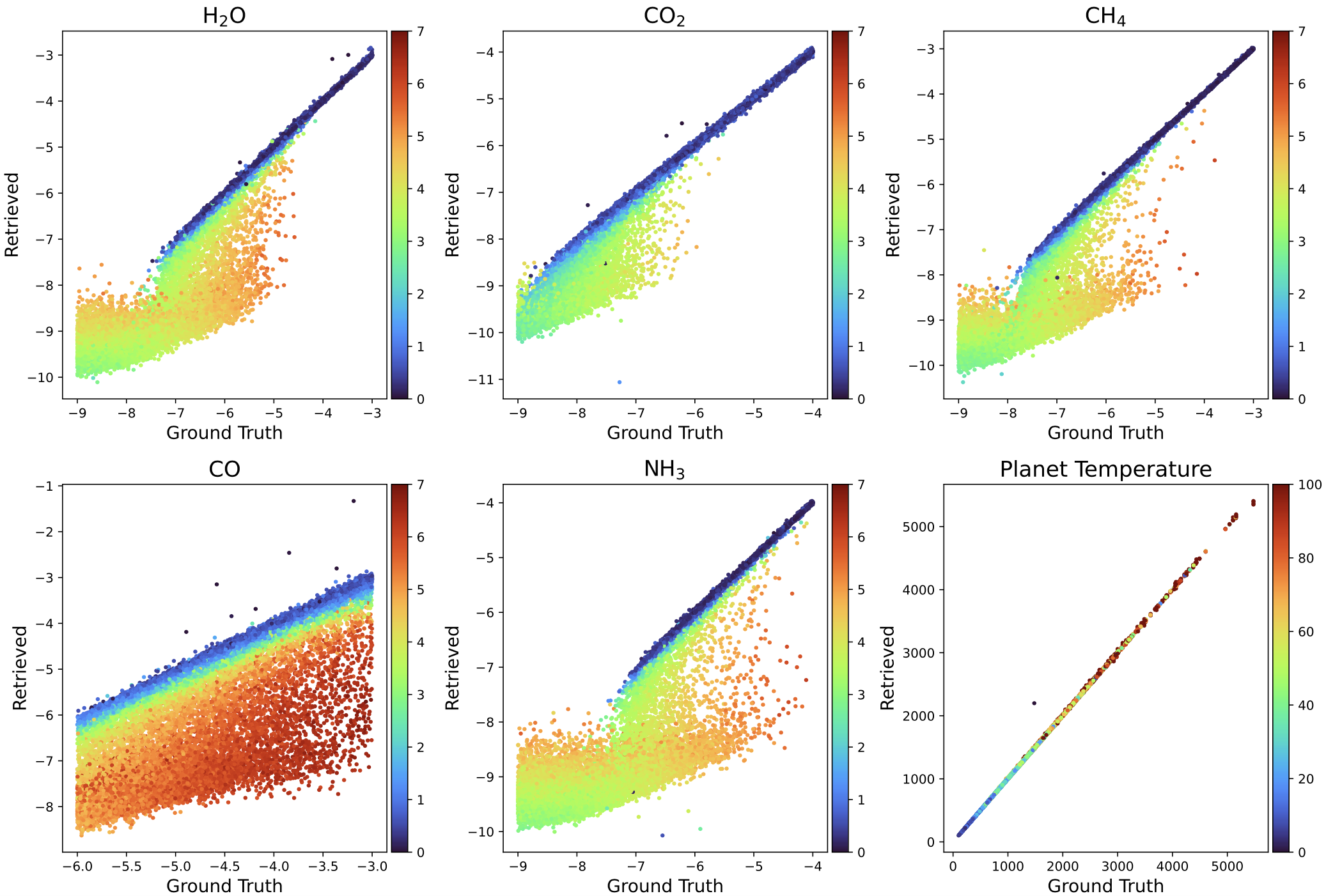}
    \caption{Comparison of the retrieved values against the input values for six different targets. Each data point represent a single instance and is colour-coded according to the respective size of the IQR. Ariel data at Tier-2 resolution is able to place tight constrain on the temperature and most molecules down to a certain abundance. Beyond that, the retrieved values starts to deviate from the diagonal and becomes less constrained, highlighting the limitations of the telescope.}
    \label{fig:PVT}
\end{figure*}

Next we will look at results from atmospheric retrieval. The quality of the retrieved product is closely related to the information content of individual spectrum, which is a function of the wavelength coverage, size of the spectral bin, observational uncertainties and the abundance of the molecule. Figure \ref{fig:PVT} compares the retrieval results against the input values of the 6 targets of interest (H$_2$O, CO$_2$, CH$_4$, CO, NH$_3$, Temperature). Each data point in every subplot represents a single spectrum and is colored in accordance to the size of the inter-quartile range (IQR)\footnote{Here we define IQR as the difference between the 84th and the 16th percentile.}. Points lying along the diagonal line - those that are retrieved correctly - tend to have tighter constraint, while points that deviate from the diagonal line tend to entail larger uncertainties. For most gases there is a transition region where molecules at certain abundance level starts to depart from the diagonal line. The extent and onset of the transition region is a function of the instrument specification (e.g. its detection limits), the composition of the atmosphere and the strength of the molecular absorption. \citet{Changeat_2020} pioneered an initial study of this transition region and derived the detection limit for each gas based on the size of the errorbar obtained. Here, we find similar results, and the detection limits of Ariel correspond to the region where all the retrieved values from Figure \ref{fig:PVT} deviate from the diagonal line (associated with colors from green to red).

Appendix \ref{sec:overview_continued} continues our discussion into other aspects of the data product.

\subsection{Structure of the ABC Database}
The database contains 2 levels of data product, the first level is for general use and the second level is designed specifically for the competition. We will describe each level below:
\subsubsection{Level 1: Cleaned Data}
Level 1 contains data products for general use. As TauREx\,3 performs forward modelling and retrieval on a planet-by-planet basis. The data is pre-processed to provide an unified structure for effective data navigation and a foundation for further processing. Below is the list of operations we performed:
\begin{enumerate}
    \item Removed any spectra with \texttt{NaN} values.
    \item Removed spectra with transit depth larger than 0.1 in any wavelength bins. 
    \item Removed spectra with transit depth smaller than $1\times10^{-8}$ in any wavelength bins. 
    \item Standardised units and data formats.
    \item Extracted all Stellar, Planetary and Instrumental metadata.
    \item Combined all instances into a single, unified file. 
\end{enumerate}
Level 1 data is organised into \texttt{all\_data.csv}, \texttt{observations.hdf5} and \texttt{all\_target.hdf5}. \texttt{all\_data.csv} contains information on the planetary system and the input values for the generation process, \texttt{observations.hdf5} contains information on individual observations and \texttt{all\_target.hdf5} contains the corresponding retrieval results (posterior distributions of each atmospheric targets). In total, there are 105,887 planet instances, 25\% of them (26,109) has complementary retrievals from Nested Sampling.

\subsubsection{Level 2: Curated Data for model training.}
The following section is designed for statistical model training. In order to allow for the broadest possible participation and minimise the overhead for non-field experts, we pre-processed the dataset with our domain knowledge so that the end product is ready for model development. At the same time, we have tailored the train/test split procedure in order to allow a diverse array of solutions and research directions. Here we outlined the list of operations we performed:
\begin{enumerate}
    \item Removed data with less than 1500 points in the tracedata. This is to allow for more accurate comparison. 
    \item Removed un-informative and duplicated astrophysical and instrumental features \footnote{ including star\_magnitudeK, star\_metallicity, star\_type, planet\_type, star\_type, star\_mass\_kg, star\_radius\_m, planet\_albedo, planet\_impact\_param, planet\_mass\_kg, planet\_radius\_m, planet\_transit\_time, instrument\_nobs.}.
    \item Split data into training and test sets (more details in Appendix \ref{sec:level2}).
\end{enumerate}
After performing the above operations, the training data has 91,392 planet instances with 21,988 of them has complementary retrievals results. The test data has 2,997 instances, all of which are complemented with retrieval results. There is a notable difference in terms of the volume of data between Level 1 and Level 2 data \textit{due to the pre-processing step and train/test split}. We have devoted a section in Appendix \ref{sec:level2} to describe the Level 2 data in details. 

\subsection{Additional resources}

Published along with the database, we provide a series of complementary resources. In particular the database is provided with a Jupyter Notebook describing the data structure, how to load the dataset, and demonstrating its main characteristics. We also include a dedicated \emph{TauREx\,3} tutorial for those eager to learn the practical aspects of building forward models and performing atmospheric retrievals. All those resources are available under the same link as the database.

\section{Open challenges}

With the constructed dataset, we intended to accelerate and incentivise dedicated efforts to tackle a number of open challenges common to both the exoplanet field and the Machine Learning field.

\subsection{Fast and Accurate Bayesian Inference}
One of the aims of the database is to enable the development of advanced inference methods that are 1. able to produce posterior distributions, but at the same time, will not require as much computational resources compared to conventional sampling based methods.
This activity is proposed as part of the goal of the NeurIPS 2022 competition with simplified atmospheric cases and has already proven very successful (Yip et al. in prep)
\subsection{Estimating and Mitigating The Effect of Data Shifts }
Machine Learning models are prone to potential performance degradation when the incoming data is different from the training distribution. This phenomenon is commonly known as data shifts \citep[e.g.][]{lu2018learning,BAYRAM2022108632}.

Any ML application to the study of exoplanetary atmosphere are likely to experience data shifts. Most ML models in the literature are currently limited to simulation-based inference as the amount of actual spectroscopic observations are fall short for model training, which has to be supplemented by simulations. The discrepancy between our simplistic atmospheric models and the actual atmosphere means that data shift is inevitable \citep[][]{humphrey2022machine}.

To emulate this situation, the test set in level 2 data are specifically designed to include chemical equilibrium forward models for which the provided ground truth from atmospheric retrievals assumed free chemistry. In some cases, clouds are included in the forward model to force degenerate behaviours in the test set \citep{Line_2016_clouds, Pinhas_2017, Mai_2019, Barstow_2020_clouds, Changeat_2021_k218, Mukherjee_2021}. Those offsets between training and test sets were voluntarily introduced to evaluate whether the performance of ML solutions remain robust and consistent under `unseen' distributions (this is typically the case in real life since we know little about real exo-atmospheres) and if they had correctly learned to faithfully reproduce the Bayesian retrieval technique.

\subsection{Adaptation to Other Atmospheric Assumptions}
Atmospheric models are physical models built on varying level of complexities and modelling assumptions. ML models, however, are trained to optimise their performance w.r.t. the provided training set/ training assumptions. In this dataset we have included forward models built from two different modeling assumptions, Simple chemsitry and Equilibrium Chemistry. It remains an open question as to how easy one can `switch' from one model assumption to another. In terms of ML terminolgy, this kind of learning falls under the umbrella of transfer learning/domain adaptation, where one strive to adopt to from source domain (original training set) to the target domain with limited number of training examples \citep[]{Wilson2020}. 

\subsection{Benchmarking Different Retrieval Techniques}
The built dataset can be used for more traditional code comparisons. The {\it TauREx} retrieval code was rigorously benchmarked against other established codes \citep{Barstow_2020, Barstow_2022}. With this dataset, the exoplanet community now has access to a wide range of well-referenced forward models and retrieval runs that can be used for standard benchmarking of atmospheric models (e.g forward models) and a diverse array of retrieval techniques \cite[e.g MCMC, Nested Sampling, Normalizing Flows: ][]{Foreman_2013, Feroz_2009, Buchner_2021, Yip_2022}.

\section{Future Expansion of ABC Database}
The database currently builds on highly simplified atmospheric model assumptions (constant or equilibrium chemistry, isothermal temperature, clear atmosphere). This is done to 1.)gauge the success of such initiatives and 2.) provide a rich dataset to complete the required task.

Future iterations could explore more complex atmospheres with much more limited amount of training examples. This is because, as more complexity is embeded into the model (for instance GCMs, complex chemistry, stellar activity effects), the computation of a single sample can take months. In this instance traditional parameter sampling is not an option, and faster AI accelerated techniques will be required. We therefore plan to further extend this database over the coming years and provide new training / test sets to develop both exoplanet and ML activities. For example, future instances of this database could feature: \\
- James Webb Space Telescope and Hubble Space Telescope complementary datasets: This would allow to develop telescope-independent ML techniques and evaluate information content between the different datasets. \\
- Other classes of exoplanets: The current set focuses on gaseous exoplanets. Future data releases could include small rocky exoplanets with secondary atmospheres, or water worlds. \\
- More complex processes: Alternative chemical model \cite[with more complete species sets, with dis-equilibrium processes: ][]{Stock_2018, Woitke_2018, Venot_2020, Al-Refaie_2022} could be provided to study retrieval biases and develop chemistry robust ML methods. Similarly, complementary sets could include stellar activity, for which the relevance of AI methods has already been shown \citep{Nikolaou_2020}, or even complex cloud models \citep{Ackerman_2001, Kawashima_2018, Gao_2020, Ma_2022}. \\
- More complex models: Eclipse observations or phase-curve observations produced using Global Climate Model could be included. This would allow to extend this to new observations as well as studying three-dimensional effects \citep{Cho_2003, Showman_2010, cho_2015, Rauscher_2008, Caldas_2019, Skinner_2021_modons, Komacek_2020, Dobbs-Dixon_2010} and to develop fast recovery techniques for phase-curve data. Current approaches to retrieve phase-curve data are limited by computational resources \citep{Irwin_2020, Feng_2020, Changeat_2021, Cubillos_2021, Changeat_2022_w103, Chubb_2022_phase} and can require up to 10 million samples (e.g weeks on HPC facilities) to fully explore the parameter space of solution with Hubble data \citep{Changeat_2021}.  \\

\section{Conclusions}

We present here the publicly available ABC Database (\url{https://doi.org/10.5281/zenodo.6770103}), an database of atmospheric forward and inverse models dedicated to the development of Machine Learning approaches in the field of exoplanets. In this paper, we introduces, for a non expert community, the basic physical and chemical processes involved in the creation of such database, describing the utilised tools\footnote{The main simulation code, TauREx\,3, is open-source and publicly available at: \url{https://github.com/ucl-exoplanets/TauREx3_public}}, and clearly stating the adopted hypothesis. The constructed set includes about 105,887 forward models and 26,109 atmospheric retrievals from conventional sampling techniques, and should serve as a community asset to explore novel techniques to solve the inverse problem of retrieving chemical composition from spectroscopic data. This database was used to support the 3$^\text{rd}$ instalment of the Ariel Data Challenge, conducted as part of the NeurIPS Conference\footnote{\url{https://neurips.cc/Conferences/2022/CompetitionTrack}}, which led to new innovative ML-based solutions to infer posterior distributions from Ariel spectra.  With this effort, and with future updates of this permanent database, we hope to facilitate the development and adoption of ML solutions to a pressing issue for the next -generation of space telescopes.

\section*{Acknowledgements}

This project has received funding from the European Research Council (ERC) under the European Union's Horizon 2020 research and innovation programme (grant agreement No 758892, ExoAI), from the Science and Technology Funding Council grants ST/S002634/1 and ST/T001836/1 and from the UK Space Agency grant ST/W00254X/1. Quentin Changeat is funded by the European Space Agency under the 2022 ESA Research Fellowship Program. The author thanks Ingo P. Waldmann, Giovanna Tinetti and Ahmed F. Al-Refaie for their useful recommendations and discussions. The authors wish to thank two anonymous referees for their useful comments.

This work utilised resources provided by the Cambridge Service for Data Driven Discovery (CSD3) operated by the University of Cambridge Research Computing Service (www.csd3.cam.ac.uk), provided by Dell EMC and Intel using Tier-2 funding from the Engineering and Physical Sciences Research Council (capital grant EP/P020259/1), and DiRAC funding from the Science and Technology Facilities Council (www.dirac.ac.uk).

\section*{Data Availability}

he data underlying this article are available as a Zenodo Digital Repository, at \url{https://doi.org/10.5281/zenodo.6770103}.



\bibliographystyle{rasti}
\bibliography{main} 




\appendix

\clearpage

\section{Introduction to atmospheric studies of exoplanets}
\label{A0:intro}

This section provides a summary of the domain knowledge required to properly exploit the ABC Database. It is written as an introduction for non-exoplanet audience.

\subsection{Observations of transiting exoplanets}

\begin{figure}
 \includegraphics[width=\columnwidth]{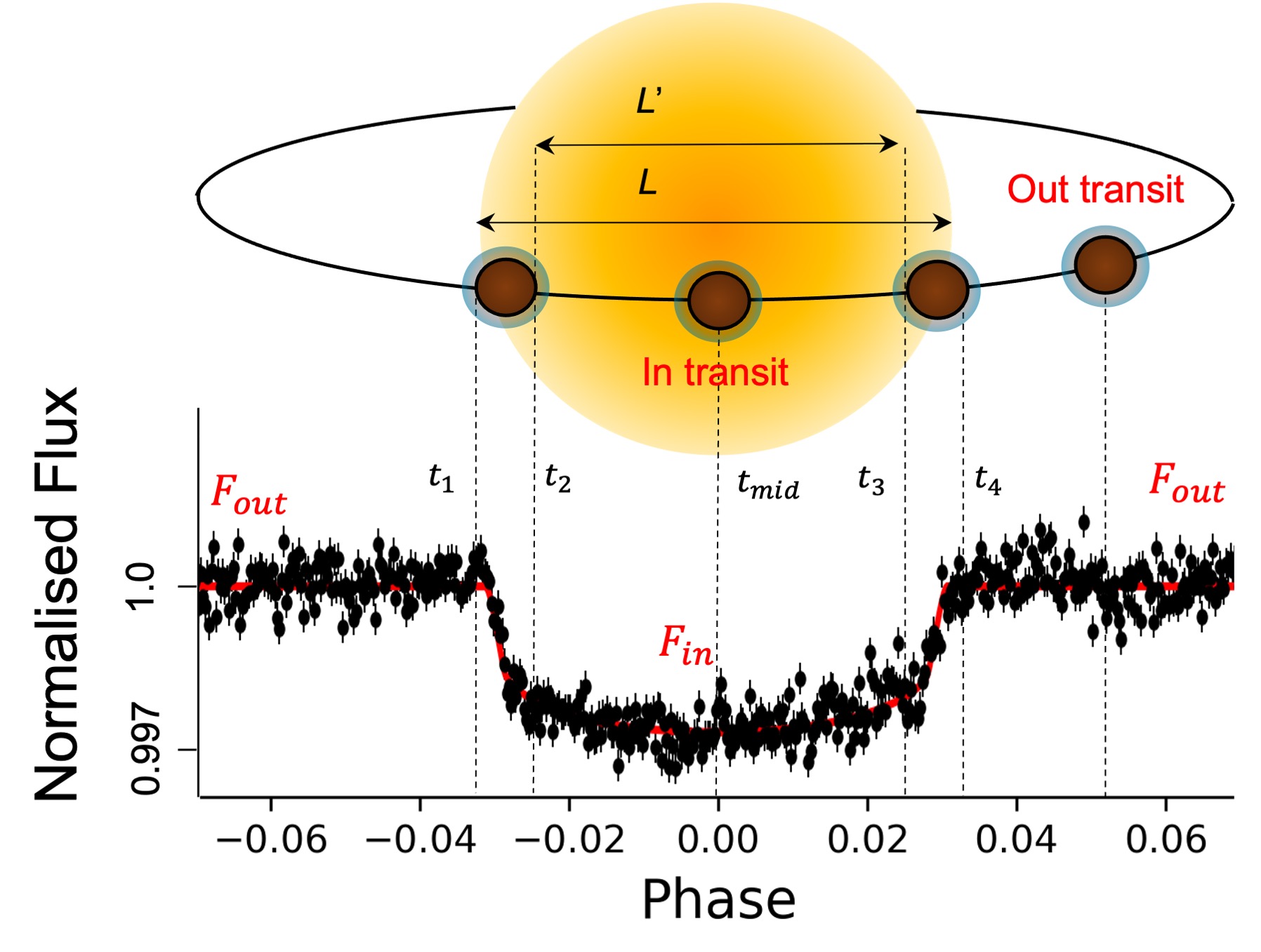}
 \caption{Diagram of an observation of the transiting exoplanet KELT-11\,b (top) and the corresponding normalised flux from a real observation, also called a light-curve (bottom). The phase, which labels the x-axis, is the position of the planet in its orbit with 0 (by convention) being the middle of the transit ($t_{mid}$). The transit starts at the event $t_1$ and finishes at the event $t_4$, spanning the transit duration $t_{14}$. The transit depth ($\Delta$) is the observed normalised flux between in and out of transit situations. The observation is adapted from \citet{Changeat_2020_k11}.}
 \label{fig:transit_obs}
\end{figure}

Exoplanets are detected using various methods, but the two most popular techniques used today are radial velocity and transit. In particular, transit is an indirect technique, which relies on monitoring the host star’s variations in brightness. A transit event occurs when the planet passes in front of the star, blocking a fraction of the light received here on Earth. Transit events can be observed, thus revealing the presence of the planet and its important properties, such as radius. A typical transit observation is described, along with the relevant quantities, in Figure \ref{fig:transit_obs}. Transit events are periodic, so they can easily be disentangled from other astrophysical sources of noise (stellar variations\footnote{for example, stars' brightness could vary from time to time}, instrument systematics and observing conditions) when long term monitoring is employed. 

For most observatories, absolute measurements are challenging. This is especially true when the required precision is high, as it is the case for exoplanets. As such, for exoplanets, we prefer to rely on differential quantities such as transit depth ($\Delta$). The transit depth is the normalised difference between the flux received from the star when the planet is out-of-transit ($F_{out}$) and when the planet is in-transit ($F_{in}$). 

\begin{equation}
    \Delta = \frac{F_{out} - F_{in}}{F_{out}} = \left(\frac{R_p}{R_s}\right)^2,
\end{equation}
where $R_p$ is the radius of the planet and $R_s$ is the radius of the star.

To first order, to account for the contribution of an atmosphere, one can simply replace the planetary radius $R_p$ by $R_p + h$, where  $h$ is the effective size of the atmosphere. Neglecting second order terms, this gives:

\begin{equation}\label{eq:deltatransit}
    \Delta = \left(\frac{R_p}{R_s}\right)^2 + \frac{2 R_p h}{R_s^2}.
\end{equation}

Now, crudely, the size of the atmosphere depends on the atmospheric scale height $H$ such that $h = N H$, where $N$ is a scaling factor encoding information regarding the atmospheric compositions. The scale height is defined as:

\begin{equation}
    H = \dfrac{k_b T}{\mu g},
\end{equation}
where $k_b$ is the Boltzmann constant, $T$ is the temperature, $\mu$ is the mean molecular weight and $g$ is the gravity. 

From those simple expressions, which here serve an illustrative purpose and are an oversimplification of the model used to build the ABC Database, we can deduce some standard behaviours of atmospheric properties. First, to extract information on the planet and its atmosphere, we will always require some knowledge of the host star. This is because the planet is not observed directly and the observed quantities ($\Delta$) are a function of the stellar parameters (here the stellar radius $R_s$). In addition, we observe that for the more massive planet (larger $g$) the contribution of the atmosphere will be diminished, as the atmosphere contracts under gravity. On the contrary, if the temperature increases the atmosphere will be inflated and thus the atmospheric signal will be larger. The chemistry of the atmosphere plays a part in the scaling factor $N$ but their relation cannot be easily deduced here. Intuitively, molecules with larger abundance tend to make the atmosphere opaque at higher altitudes, therefore increasing the apparent size of the atmosphere\footnote{we cannot observe any non-opaque (transparent) part of the atmosphere}. 

Those concepts, while useful to acquire an intuitive understanding of the behaviour of planetary atmospheres, are rather limiting and proper modelling is required to correctly interpret exo-atmospheric observations.

\subsection{Modelling exoplanet atmospheres}
\label{sec:atmosphere}

Observing exoplanetary transits at various wavelengths, meaning obtaining $\Delta$ as a function of $\lambda$, provides information about the atmospheric properties. This is because a planetary atmosphere contributes to the transit depth by absorbing the incoming stellar light (slightly) differently at different wavelengths (e.g. the atmospheric contribution is wavelength dependent). The absorption profile of the atmosphere depends on its constituents (molecular species, clouds, hazes) and properties (thermal structure). To model the observed signal as a function of wavelengths, also called a spectrum, astronomers use simplified models of the relevant processes occurring in exoplanet atmospheres. In Appendix \ref{A1:atm_physics}, we describe the mathematical formulation of one such model for the transit geometry, commonly used as a parameterised 1-D forward model. Put simply, the light from the host star is propagated through an atmosphere layer-by-layer and impacted according to the absorption of the atmosphere. In our case, the absorptions considered are absorptions by molecular species, Rayleigh scattering and Collision Induced Absorption (CIA). 

Through this process, from a parameterised one-dimensional description of an atmosphere controlled by a finite number of parameters, one can compute the theoretical spectrum of an exoplanet. This process, called forward modelling, can be made relatively fast (on the order of seconds), but due to the non-linearity of the equations and the input spectroscopic data (cross-sections) it cannot be directly inverted. In the next section, we explain how traditional techniques (e.g Bayesian sampling) are used to perform model fitting and retrieve the properties of an exo-atmosphere from its observed spectrum. 

Before explaining the use of inversion techniques, or atmospheric retrievals, applied in the context of exoplanet atmospheres, we wish to present a series of simple models to illustrate further the sensitivity analysis made in the previous section. We have created a mock planet with a non-negligible atmosphere, and we will show how changing the values of some of the model parameters affects the observation (e.g the spectrum). 

\begin{figure*}
 \includegraphics[width=1.9\columnwidth]{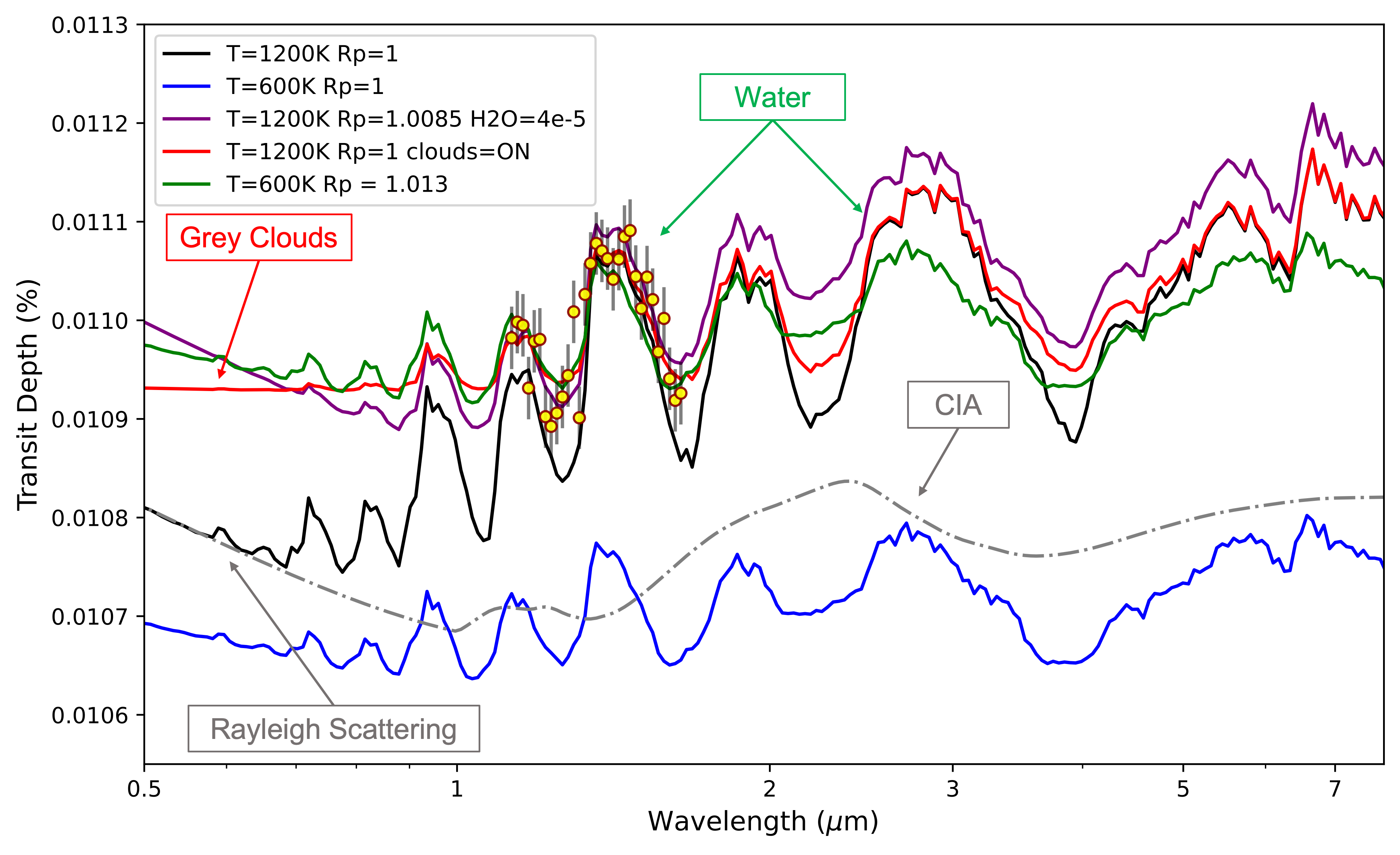}
 \caption{Spectra illustrating the sensitivity of atmospheric models to input parameters. In black: Model 1, in blue: Model 2, in purple: Model 3, in red: Model 4, and in green: Model 5. We also show in dashed grey line a model similar to Model 1 but without absorption of water, leaving only the continuum contribution of Rayleigh Scattering (short wavelengths) and CIA (long wavelengths). The red and yellow points represent a simulated observation with HST at 30 parts per million (ppm), highlighting the difficulty of constraining atmospheric properties from current data.}
 \label{fig:sensitivity}
\end{figure*}

For simplicity, we set the planet with an isothermal atmosphere, meaning the temperature of the atmosphere is constant with altitude (e.g: constant at all pressure levels) and therefore can be defined by a single parameter (T). To this atmosphere, we add a single trace molecule (H$_2$O) defined by its absolute abundance in volume (volume mixing ratio), and we fill the rest remaining atmosphere with hydrogen and helium in standard solar ratios (H$_2$/He = 0.17)\footnote{The trace molecules (like H$_2$O only accounts for a very tiny portion of the atmospheric composition, the rest is filled by gases like Hydrogen and Helium, this kind of atmosphere is also known as Primary Atmosphere (For instance, Jupiter has a Primary Atmosphere) as opposed to Secondary Atmosphere, which is principally made of heavier elements (For instance, Earth has a Secondary Atmosphere). }. On top of the molecular absorption from water vapour, we also consider three additional absorption processes: Collision Induced Absorption (CIA), Rayleigh Scattering and Grey Clouds (not considered in this version of the ABC database). Equipped with this model, we set the following cases for which the spectra are available in Figure \ref{fig:sensitivity}: \\

$\bullet$ Case 1 (black): Planetetary radius R$_p$ = 1.0 R$_J$, Temperature T=1200K, Mixing ratio of H$_2$O = 10$^{-3}$ and no clouds. \\

$\bullet$ Case 2 (blue): Same as Case 1 but the temperature is decrease to T= 500K. \\

$\bullet$ Case 3 (purple): Same as Case 1 but the water content is decrease to H$_2$O = 10$^{-5}$ while the planetary radius is increased to R$_p$ = 1.0085. \\

$\bullet$ Case 4 (red): Same as Case 1 but with clouds (cloud top pressure is set at 0.01 bar). \\

$\bullet$ Case 5 (green): Same as Case 2 but with an increased planetary radius to R$_p$ = 1.013 R$_J$. \\

From those specifically designed case, one can compare Case 1 and 2, for which only the temperature is changed. As a consequence of this change, the size of the atmosphere is decreased as explained in Section 2.1, and the atmospheric features are smaller, bringing the whole spectrum down. In this case, distinguishing between Case 1 and Case 2 would be relatively easy. For Cases 3, 4 and 5, however, the story can be a little more complicated as multiple parameters are changed, but those can be used to highlight degeneracies typically encountered in the interpretation of exoplanet spectra, therefore justifying the need for more sophisticated atmospheric retrieval techniques. 

For those cases, the spectral features are reduced compared to Case 1, but they appear much closer in the 1$\mu$m  to 2$\mu$m wavelength range. This is because Case 3 has less water compared to Case 1, which we expect to decrease the spectral features but thanks to the slightly larger radius, the spectrum is brought back to a similar level. Case 4 has opaque clouds, which "cuts" the spectral features above a certain pressure level, making it look exactly like Case 3 in the 1$\mu$m to 2$\mu$m range. Finally, Case 5 has a lower temperature (600K) and is brought back to the same level by an increased in radius. With current telescopes, such as HST, the wavelength coverage is relatively small. One typical instrument onboard HST is the Wide Field Camera 3 with its G141 Grism, which has a wavelength coverage from 1.1$\mu$m to 1.6$\mu$m and reaches errors of the order 30ppm\footnote{parts per million, 10$^{-6}$}. Highlighting a typical observation with HST on the same figure, we show how difficult it would be to distinguish between Cases 3, 4 and 5. This highlight the requirement to next-generation space telescopes such as Ariel to constrain atmospheric properties.

\subsection{Solving the inverse problem for exo-atmospheres}
\begin{figure}
    \centering
    \includegraphics[width=0.7\columnwidth]{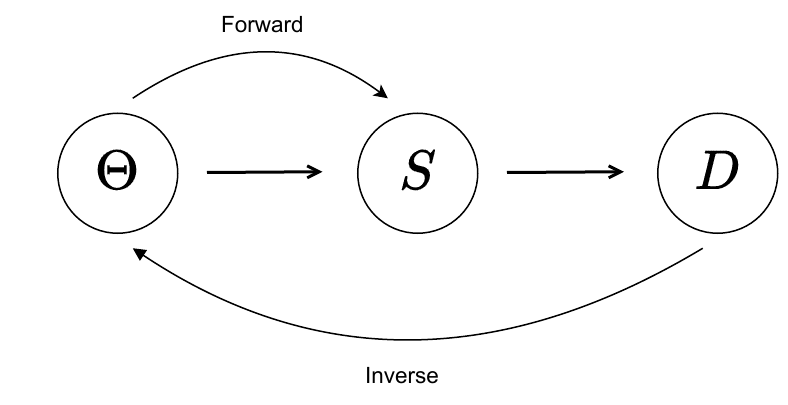}
    \caption{Schematic of a typical inverse problem setup. The forward process produces an effect (full spectrum, $S$) from a hidden cause (e.g. atmospheric parameters, $\Theta$). However, the full effect is often unavailable to the observer due to loss in information (such as instrument systematics, limitation in spectroscopic coverage etc. ). Instead, observers can only receive the partial effect (or otherwise known as the observation, $D$). The aim of the inverse problem is to recover the hidden cause that produces them in the first place. }
    \label{fig:inverse}
\end{figure}
The study of exo-atmosphere relies on spectroscopic observations to infer fundamental atmospheric properties that cannot be directly observed. This kind of problem is broadly described as the inverse problem \citep[][]{Potthast_2006}, where one tries to uncover the cause (atmospheric properties) from the effect (observations). However, more often than not, the full effect is seldom observed, instead, observers often received a corrupted form of the effect, which is the observations. In terms of exoplanetary spectra, there are several sources of corruption, such as the presence of noise and limited spectroscopic coverage. The loss in information often means that the inverse mapping function $\mathcal{M}^{-1}$ is unknown and may no longer be uniquely defined, which generally give rise to more than one plausible causes, also known as model degeneracy (see section above). In some extreme cases, severe loss in information (like extremely low S/N observations) effectively means that the cause may no longer be recoverable. See Figure \ref{fig:inverse} for a typical setup of an inverse problem.

Our goal is to estimate the set of parameters $\Theta$ that best explains the observed spectrum $D$ under a given atmospheric model $\mathcal{M}$. There are different ways to approach this "atmospheric retrieval" problem, but most of them involve a forward model (which includes our atmospheric assumptions) and an optimizer. Here we will briefly describe the problem in terms of Bayesian framework, for a more detailed discussion on Bayesian Statistics, one can refer to \cite{skilling_2006, Feroz_2009,Foreman_2013, MCMC,Trotta, Speagle_2020, Buchner_2021} for more information. 

Our goal is to find the conditional distribution of the model parameters given the observation, also known as the posterior distribution ($P(\Theta|D, \mathcal{M})$) in Bayes' Theorem \citep[][]{Bayes}. The posterior distribution can be computed via the following formulation:
\begin{equation}
    P(\Theta|D, \mathcal{M}) = \frac{P(D|\Theta, \mathcal{M})P(\Theta)}{P(D)}
    \label{eqn:bayes}
\end{equation}
where $P(D|\Theta,\mathcal{M})$ represents the likelihood function under a given model, $P(\Theta)$ represents the prior and $P(D)$ represents the normalising constant, or the Bayesian Evidence. 

The (log-)\,Gaussian likelihood function is commonly used to compare the observation $D$ with the output from the forward model $\mathcal{M}$, i.e.
\begin{align}
    \mathbb{E}[\text{log} P(D|\Theta,\mathcal{M}] &= \mathbb{E}[\text{log}(\mathcal{N}(D|S, \sigma)] \\
     &= \mathbb{E}\left[\text{log}\left(\frac{1}{\sqrt{2\pi\sigma^2}} \text{exp}\left(-\frac{1}{2}\frac{(D - S)^2}{\sigma^2}\right)\right)\right]
    \label{eqn:llh}
\end{align}
where $S$ is a simulated spectrum generated using the forward model $\mathcal{M}$ and $\sigma$ represents the observation uncertainty/noise. Thanks to the forward model $\mathcal{M}$, we have an unique mapping from a set of parameters to a simulated spectrum $S$ such that:

\begin{equation}
    S = \mathcal{M}(\Theta).
\end{equation}

The relation between the observed spectrum $D$ and the simulated spectrum $S$ is:

\begin{equation}
    D \approx S + \epsilon,
\end{equation}
where $\epsilon = N(0, \sigma^2)$. The approximation sign reflects the fact that the model remains an approximation of the real phenomena.

As for the prior function $P(\Theta)$, it represents our prior belief on the distribution of the random variables. With limited knowledge on the exo-atmosphere, the community always opt for an uninformative prior ( also known as an uniform prior).

Unfortunately, in most cases Equation \ref{eqn:bayes} cannot be computed analytically. The main reason lies with the Bayesian Evidence, $P(D) = \int P(D, \Theta) d\Theta$, the integral demands evaluation of the probability for every possible combinations, which makes the quantity intractable for any meaning cases.

A common strategy is to sample the parameter space, and use the distribution of the samples to compute the maximum likelihood estimation (MLE) and the Bayesian evidence. There are many optimizing strategies available, these includes grid sampling, optimal estimation, Markov Chain Monte Carlo models (MCMC) and Nested Sampling amongst others. Those are however computationally intensive and require evaluation of millions of forward models, 

There have been efforts from the ML community to develop scalable sampling algorithms. Stochastic Gradient MCMC (SG-MCMC) is a popular class of algorithms that utilises data sub-sampling techniques to reduce computational time to construct the chain \citep[][]{Welling_2011, Ma2015ACR,baker_control_2019, Nemeth_2019}. Stochastic Gradient Descent(SGD)'s link to approximate Bayesian inference has prompted further investigation into its statistical properties \citep[][]{Mandt_SGD, chen2016, Xing_2018}, it has since been shown that SGD with constant step size (Constant-SGD) can approximate Bayesian Posterior Distribution. Other algorithms, such as Hamiltonian Monte Carlo (HMC), incorporated information on the gradient within the proposal to improve the sampling efficiency \citep[][]{Neal2011,Homan2014}. \cite{Chen2014} introduced SG-HMC, a fusion between SG-MCMC and HMC, to provide further speed up to the algorithm.

Other approaches focuses on architectural design or post-processing techniques to incorporate elements of Bayesian Inference, such as Dropout \citep[][]{Gal_2015}, Neural Network Ensembles \citep[][]{Lakshminarayanan2017, Pearce2018, Cobb2019}, SWA-Gaussian \citep[SWAG][]{Maddox2019}, KF-Laplace \citep[][]{ritter2018scalable}, temperature-scaling \citep[][]{Guo2017}.

The availability of many state-of-the-art algorithms prompts the need to benchmark their performances under different datasets and scenarios \citep[][]{Yao2019,Izmailov2021}. Aligned with this objective, the aim of this database and the machine learning challenge is to leverage recent developments in scalable Bayesian Inference and identify potential solutions forward.

\section{Atmospheric Transmission Model in \emph{TauREx}}
\label{A1:atm_physics}

\begin{figure*}
 \includegraphics[width=\textwidth]{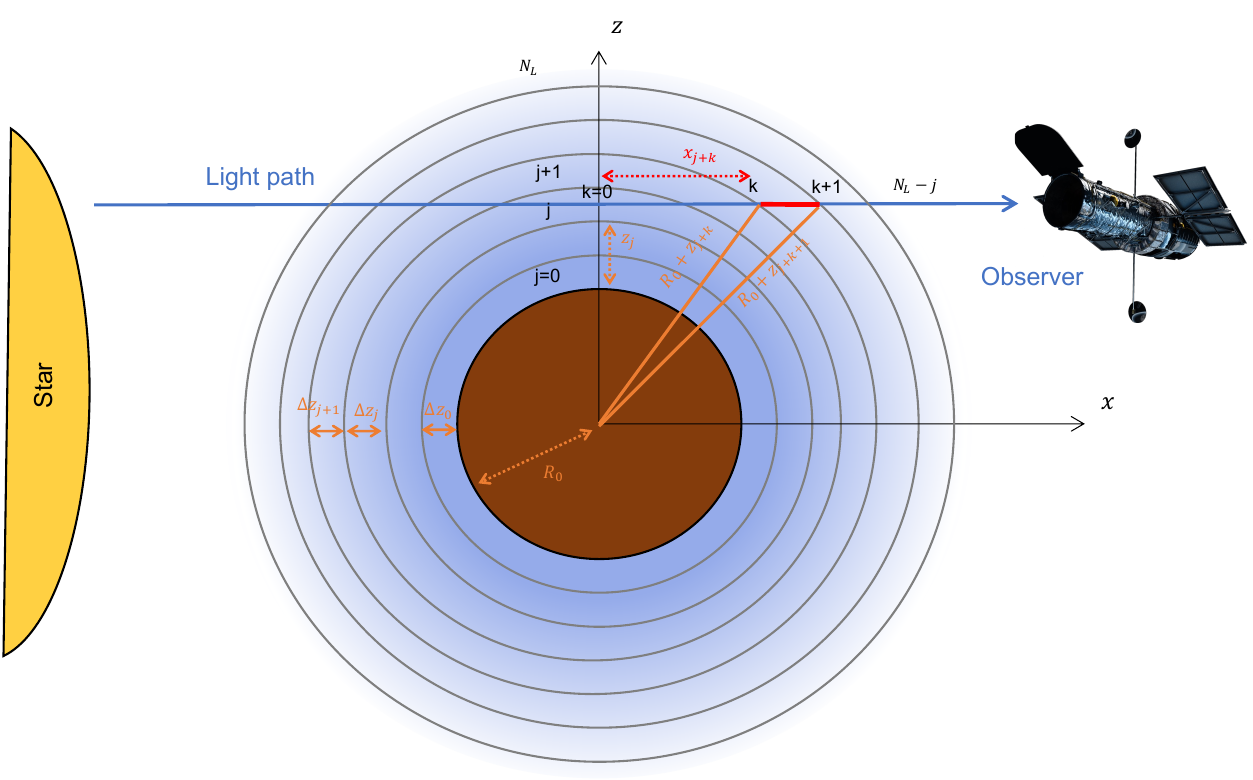}
 \caption{Illustration of the transmission of stellar radiation (left side) through an exoplanet atmosphere (transit) towards an observer (right side). R$_0$ is the reference radius at which the atmosphere becomes fully opaque. A light ray at altitude $z$ propagates along the line of sight $x$. The atmosphere is separated in N$_{L}$ layers for size $\Delta_z$, which are labelled by the index $l = j + k$, where $j$ refers to the $z$ component and $k$ to the $x$ component. The discretised altitude z$_l$ corresponds to the altitude at the lower boundary of the layer $l$.}
 \label{fig:transit_atmo_model}
\end{figure*}

In this section, we describe the simplified transit (forward) model used in the code \emph{TauREx\,3}. The atmosphere of the planet is separated in N$_{L}$ homogeneous layers following a one-dimensional plane-parallel geometry (see Figure \ref{fig:transit_atmo_model}). The light-rays from the host star are propagated through the atmospheric layers, being impacted by extinction processes (absorption and scattering) at the different wavelengths ($\lambda$).

The normalised differential flux ($\Delta_{\lambda}$) or the transit depth at wavelength $\lambda$  reaching the observer is simply the ratio of the surface area of the planet to the host star, which can be further simplified to the planet-to-star radius squared:

\begin{equation}\label{eq:transit_depth_eqn}
    \Delta_{\lambda} = \frac{F_{out,\lambda} - F_{in,\lambda}}{F_{out,\lambda}} = \left(\frac{\pi R_p(\lambda)}{\pi R_s}\right)^2 =  \left(\frac{R_p(\lambda)}{R_s}\right)^2,
\end{equation}
where $F_{out,\lambda}$ is the total flux received from the system out-of-transit, $F_{in,\lambda}$ is the total flux received in-transit, $R_p(\lambda)$ is the wavelength-dependent radius which includes the atmospheric contribution and $R_s$ is the stellar radius. In our case, the atmospheric contribution consists in the absorption of the star light from the atmosphere (e.g. we do not include scattering processes), which follows the Beer-Lambert law.

The wavelength-dependent contribution of the atmosphere starts at the surface labelled $R_0$. Note that for gaseous planets (e.g. without solid surface), $R_0$ is a reference radius at which we consider the atmosphere is fully opaque at all wavelengths. We get:

\begin{equation}\label{eq:contribution_eq}
   \pi R_p(\lambda)^2 = C_{sur} + C_{atm} = 2 \pi \int_0^{R_0} r dr + 2 \pi \int_{R_0}^{\infty} r \left(1- e^{-\tau(r,\lambda)} \right) dr,
\end{equation}
where $C_{sur}$ is the contribution to the planet surface, $C_{atm}$ is the contribution from the atmosphere and $r$ is the radial coordinate. In most cases, the former term is assumed to be completely opaque and therefore can be simply evaluated as the surface area of the planet at radius $R_0$, the latter term involves the computation of the optical depth of the atmosphere at each layer, which summarises the contribution from various processes happening within the atmosphere.

The optical depth $\tau(r,\lambda)$ is computed along the line of sight as follow:

\begin{equation}\label{eq:optical_depth}
    \tau(r,\lambda) = 2 \int_0^{x_f} \sum_i^{N_{G}} \chi_i (r') \rho(r') \sigma_i(r', \lambda) dx. 
\end{equation}
Here, $\chi_i$ is the mixing ratio (or abundance) of the $i^{th}$ species, $\rho$ the number density, and $\sigma_i$ the absorption cross-section of the $i^{th}$ species. The number of gases is noted $N_G$. The variable $x_f$ is the maximum distance considered for the numerical integration.

Considering the one-dimensional geometry, the integration of $\tau$ along the $x$ axis can be decomposed in unit elements $\tau(j,k)$, where $j$ represents the $y$ axis indexes and $k$ are the indexes along the $x$ axis. Physical quantities (e.g: the altitude $z$, the mixing ratio $\chi$) defined at a layer $l$ can then be related to the $j$,$k$ indexes using $l = j + k$ and noting that $k$ can only span the values from $j$ to N$_{L}$. These are indexed with an additional subscript, for instance $\chi_{i,l}$ is the mixing ratio of the $i^{th}$ species at layer $l$.

It follows that the unit path integral, labelled $\Delta x_{(j,k)}$ and identified by the red element in Figure \ref{fig:transit_atmo_model}, can be expressed as:

\begin{equation}
\begin{split}
    \Delta x_{(j,k)} = & \sqrt{ \left(R_0 + z_{j+k+1} \right)^2 - \left( R_0 + z_j + \frac{\Delta z_j}{2}\right)^2} \\
    & - \sqrt{ \left(R_0 + z_{j+k} \right)^2 - \left( R_0 + z_j + \frac{\Delta z_j}{2}\right)^2},
\end{split}
\end{equation}
where $z_l$ is the altitude at layer $l$ and $\Delta z_l$ is the changes in altitude at layer $l$.

Since the layer are equally spaced in log-pressure we also have:

\begin{equation}
    \Delta z_l = - H_l \text{log} \left(  \dfrac{P_{l+1}}{P_l}\right),
\end{equation}
where $H_l$ is the scale height at layer $l$ and P$_l$ is the pressure at layer $l$.

Expressing the optical layer element as

\begin{equation}
    \tau_{(j,k)} = \sum_i^{N_{G}} \chi_{i, j+k} \rho_{j+k} \sigma_{i, j+k}(\lambda) \Delta x_(j,k),
\end{equation}

one gets the final contribution for the  atmosphere as:

\begin{equation}
    C_{atm} = 2 \pi \mathlarger{\sum}_{j=0}^{N_{L}}  (R_0 + z_j)\left( 1 - exp\left[ -2 \sum_{k=0}^{N_{L}-j} \tau_{(j,k)}\right] \right) \Delta z_j,
\end{equation}
and the transit depth as a function of wavelengths or the transmission spectrum can be computed following equation \ref{eq:transit_depth_eqn}. In this investigation, we produced a grid of transmission spectra through a randomised and uniform grid of free parameters. 

The absorbing properties of the different molecules (H$_2$O, CO, CO$_2$, CH$_4$ and NH$_3$) and processes (Rayleigh Scattering, CIA) are encoded in the cross-sections ($\sigma_i$ in Equation \ref{eq:optical_depth}). Cross-sections are temperature, pressure and wavelength dependent, and have a highly non-linear behaviour. In most codes, including the one used here, since the computation of cross-sections is a computationally intensive and complex process, they are pre-computed in tabulated files, which are then interpolated to obtain the absorbing profile of the relevant molecules and processes at a given temperature, pressure and wavelength. 

For a more complete description of the employed code, we refer the reader to the original papers \citep{Al-Refaie_2021, Al-Refaie_2021_t31} and the NeurIPS taurex tutorial available at Zenodo: \url{https://doi.org/10.5281/zenodo.6770103}.

\section{Posterior distribution of atmospheric retrieval}
\label{A2:posteriors}

\begin{figure*}
 \includegraphics[width=\textwidth]{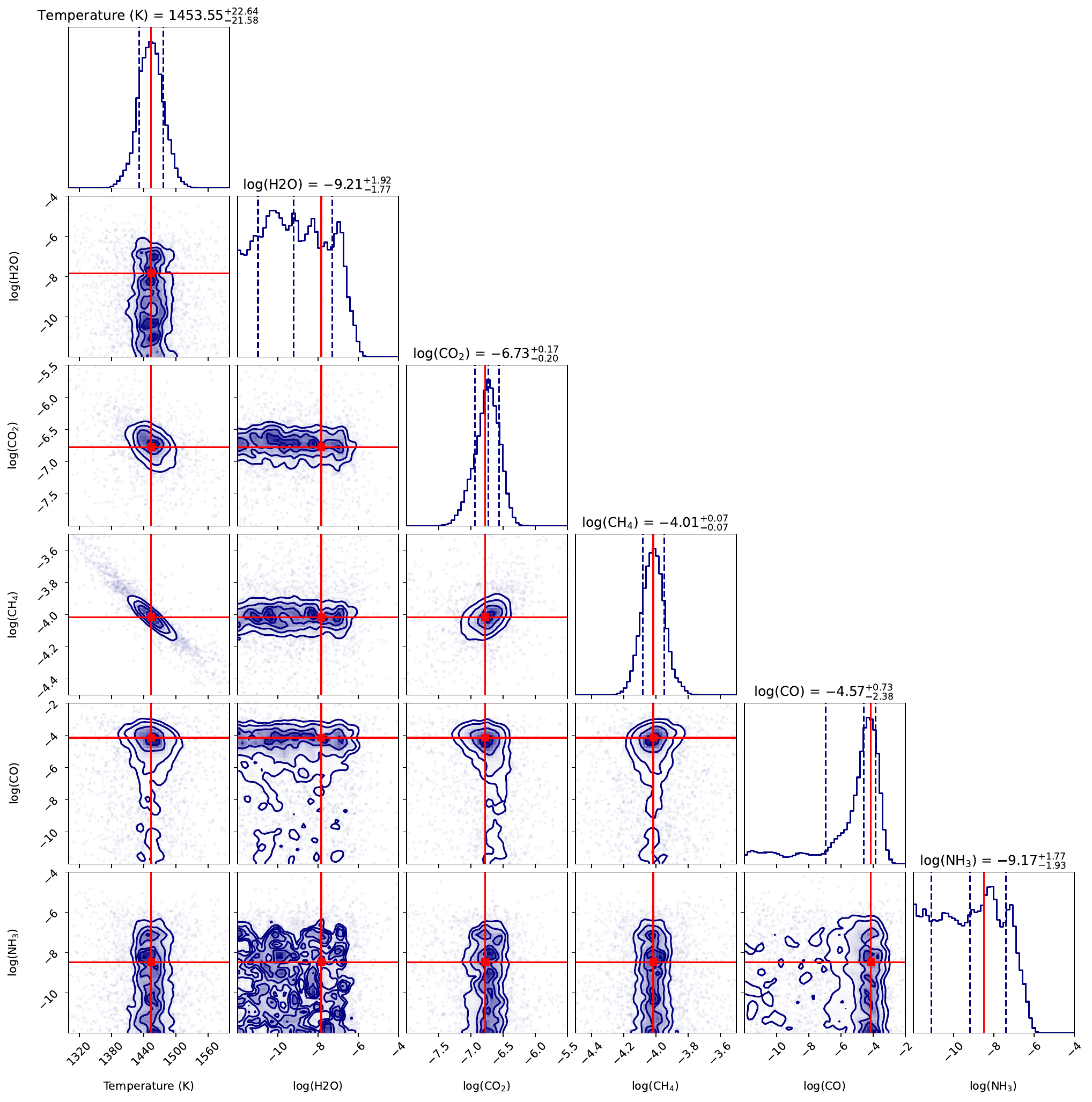}
 \caption{Example of posterior distribution obtained with \emph{TauREx\,3} on a simulated Ariel observation. This correlation map is constructed using the Nested Sampling traces and weights, with the \emph{corner} library.}
 \label{fig:posteriors_ariel}
\end{figure*}

Figure \ref{fig:posteriors_ariel} shows an example of posterior distribution resulting from a \emph{TauREx} atmospheric retrieval. This posterior distribution corresponds to the data shown in Figure \ref{fig:retrieval_ariel}. 

The posterior distribution shows the correlation between the free parameters of the model (here atmospheric temperatures and abundances of five gases). In particular, this inverse problem is challenging for ML solutions as, due to the high level of degeneracies between the parameters of interest, the exoplanet community is interested in obtaining full probability distributions rather than a unique guess. Solutions to this inverse problem would be required to: 1) correctly identify the abundances of detectable molecules (see CO$_2$, CH$_4$ and CO); 2) Characterise the correlation between parameters (see for instance the negative correlation between temperature and abundance of CH$_4$); 3) Constrain upper limits for the parameters that cannot be determined (see for instance NH$_3$ distribution); 4) Identify multimodal solutions (not shown in this example).

\section{Data Overview - continued}
\label{sec:overview_continued}
A strength of this large-scale data generation lies with the use of currently known demographics as the source of planetary candidates. Planet formation and evolution remains an actively researched area and there are contrasting theories as to how and why certain planets are more prevalent than others. By relying solely on observed planets we avoided producing fictitious planets that are otherwise impossible to form. Furthermore, the bi-model distribution of planet mass and radius contributes to the dichotomy seen in Figure \ref{fig:two_distribution}.

\begin{figure*}
    \centering
    \includegraphics[width=\textwidth]{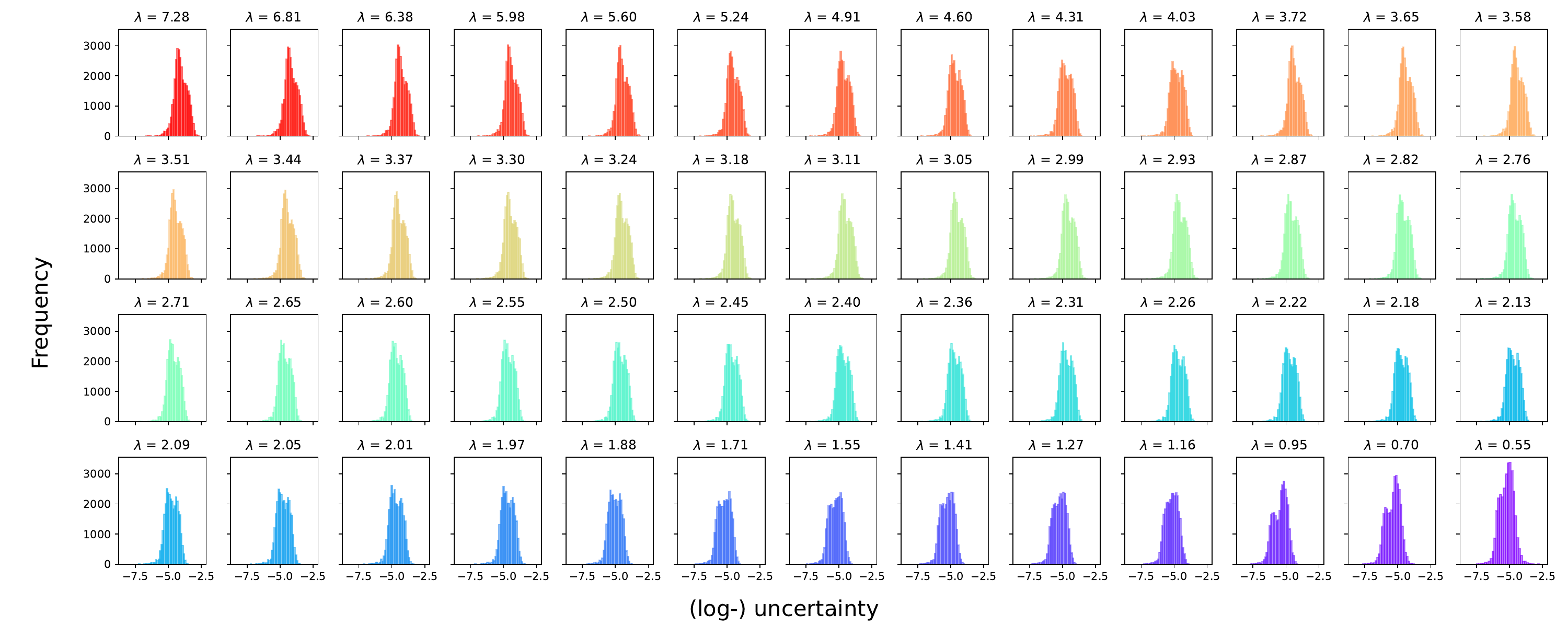}
    \caption{Distribution of (log-) uncertainty across different wavelength channels used by Ariel-Tier 2 resolution. These uncertainties are generated using ArielRad, which accounts for the different instrumentation on board Ariel, stellar properties as well as planetary properties. Since the SNR requirement of Ariel Tier-2 data is on the atmospheric signal, those distribution are approximately offset by 1 order of magnitude compared to the "Feature height" distribution in Figure \ref{fig:two_distribution}}.
    \label{fig:arielrad_wl_distribution}
\end{figure*}

Due to the extremely low S/N ratio with exoplanetary observation and non-linear instrument systematics, actual observations are usually accompanied with non-negligible measurement errors. These errors are specific to the brightness of the host star, the data reduction process, the instrument onboard and its separation from us. ArielRad, an official radiometric simulator dedicated to the Ariel Space Mission, is designed specifically to account for the aforementioned  effects and provide realistic estimation of the observational uncertainties \citep{Mugnai_2020_arielrad}. Figure \ref{fig:arielrad_wl_distribution} shows the distribution of (log-) observational uncertainties across the 52 wavelength channels. All of them displayed a non-Gaussian distribution, some even presented a bi-model distribution. There are also noticeable difference in terms of the shape and magnitude across different channels. For instance, uncertainties associated with the blue end of the spectrum tend to be smaller than the red end of the spectrum, as the blue end of the spectrum is coming from a photometer.


\section{Level 2 Data - Detailed descriptions}
\label{sec:level2}
\subsection{Structure}
Level 2 data is designed originally for the NeurIPS 2022 competition, but the data structure can be re-used for general model training. It is consisting of a training and test set. The two sets shares the same structure and the aim is to allow better readability to non-field experts.

\begin{enumerate}
    \item \texttt{AuxillaryTable.csv:} contains supplementary astrophysical parameters.
     \item \texttt{SpectralData.hdf5:} contains details of the spectroscopic observations
     \item \texttt{Ground Truth Package:} The package contains the ground truth targets for the competition. 
     \begin{enumerate}
         \item \texttt{TraceData.hdf5} records the traces of the empirical distribution obtained from Nested Sampling, it is primarily used for the Regular Track. 
         \item \texttt{QuartilesTable.csv} records the 16th, 50th and 84th  percentile of the posterior distribution, it is mainly used as a target for the Light Track.
         \item \texttt{FM\_Parameter\_Table.csv} records the model values that generates the spectra in the first place. While it could be different from the ground truth, it can be used as a soft label. 
     \end{enumerate}
\end{enumerate}

\subsection{Train-Test Split}

With our long separation from any exoplanets and limitations from current technologies, it is almost impossible to ascertain the true nature of the target exo-atmosphere. In other words, \emph{our test distribution will always be different from the training distribution, also known as domain shift in Machine Learning literature} \citep[][]{Wang_2018,Wilson2020} To reflect this limitation, the train/test split is designed to uncover solutions that can maintain their performance even under unknown situations (unseen atmospheric behaviour and/or unseen planets). 

To support this goal, we abandoned the usual practise of dividing a dataset randomly into training and test set, which tests the model's ability to generalise under a \emph{homogeneous} distribution. Instead, the test set is designed to contain In-Training Parameter Ranges (In-Range) and Out-of-Training Range Parameters (Out-Range) components. In-Range samples represent examples that come from the same distribution as the training data. Out-Range represent samples that are unseen by the model during training, this includes unseen planetary and atmospheric properties.

As a result, some of the planets are purposely removed from the training set to create \emph{unseen} planetary properties, any theoretical spectra created using from those planets are also taken away from the training, causing a slight drop in the amount of available training data, as compared to Level 1 data. We further generated 5461 spectra under equilibrium chemistry scheme \citep[][]{Agundez_2012,Agundez_2020} by assuming solar elemental ratios to create \emph{unseen} atmospheric properties. As stated above, these spectra are unseen and thus are \emph{not} included in the training set. 

The test set is stratified into 4 subsets, each representing varying degree of similarity to the training data. Table \ref{tab:sets} summarises the different configurations of the 4 subsets. Subset 1 is the most similar to the training set as all the components are In-Range , while Subset 4 is the most dissimilar as all the components are Out-Range. All of them contain roughly the same proportion ($\sim$25\% )in the test set. 

\begin{table}
\centering
\resizebox{\columnwidth}{!}{%
\begin{tabular}{lllll} 
\toprule
                        & Subset 1           & Subset 2               & Subset 3               & Subset 4               \\ \hline
Planetary Configuration & In-Range & Out-Range    & In-Range & Out-Range \\
Atmospheric Properties   & In-Range & In-Range & Out-Range     & Out-Range \\
\bottomrule
\end{tabular}
}
\label{tab:sets}
\end{table}

All examples, regardless of their initial atmospheric assumptions or planetary properties, are homogeneously retrieved using the free chemistry settings outlined in Section \ref{sec:retrieval}. By doing so our retrievals will be, on purpose, biased and will not be retrieving the input chemistry (ground truth). Participants are tasked to reproduce the same results from our biased retrievals.

The combined effect of these two changes means that any proposed solution will have to maintain reliable and consistent behaviour when exposed to distributions that are unknown and unseen to their training distribution. We explicitly did not include any spectra generated with equilibrium chemistry assumption in the training set, as a proxy of the actual situation - our atmospheric models cannot adequately describe the actual atmosphere. 

The stratification of test examples provides flexibility to future investigation. The test set can be used to test the trained model under different testing conditions, for instances, one can test their models on set 1 examples to understand the model's performance under homogeneous cases. In any cases, spectra generated with either free chemistry and/or equilibrium chemistry is available online for any interested parties to construct their own training and test set. 

The dataset for NeurIPS 2022 competition contains a similar data structure, but featured more diverse. unseen atmospheric assumptions than the one presented here. A discussion of these atmospheric assumption is outside the scope of this paper, readers are advised to refer to  \cite{Yip_2022_neurips} for a more detailed description of the respective test set for the data challenge.

\bsp	
\label{lastpage}
\end{document}